\documentclass[11pt]{article}

\usepackage{amsmath,amssymb,mathtools}
\usepackage{bm}
\usepackage{booktabs}
\usepackage{longtable}
\usepackage{graphicx}
\usepackage{algorithm}
\usepackage{algpseudocode}
\usepackage[a4paper,margin=1in]{geometry}
\usepackage{amsthm}
\usepackage{hyperref}

\numberwithin{equation}{section}
\newtheorem{proposition}{Proposition}[section]
\newcommand{\newsiamremark}[2]{\newtheorem{#1}{#2}[section]}
\newcommand{\headers}[2]{\markboth{#1}{#2}}
\newcommand{\email}[1]{\href{mailto:#1}{\texttt{#1}}}
\newenvironment{keywords}{\par\smallskip\noindent\textbf{Keywords.}\ }{\par}
\newenvironment{MSCcodes}{\par\smallskip\noindent\textbf{MSC codes.}\ }{\par}
\date{}
\newsiamremark{remark}{Remark}

\newcommand{\R}{\mathbb{R}}
\newcommand{\sym}{\operatorname{sym}}

\newcommand{\cR}{\mathcal{R}}
\newcommand{\cH}{\mathcal{H}}
\newcommand{\cI}{\mathcal{I}}

\title{Structure-Preserving Full Approximation Scheme for Real-Space Hartree--Fock Equations}

\author{
Yuejia Zhang\thanks{
  School of Mathematical Sciences,
  Fudan University, Shanghai, China
  (\email{yuejiazhang21@m.fudan.edu.cn}).}
}

\headers{Structure-Preserving FAS for Hartree--Fock}{Y. Zhang}

\begin{document}

\maketitle

\begin{abstract}
We develop a structure-preserving full approximation scheme (FAS) for the
real-space restricted Hartree--Fock equations. The method treats the orbital
equations, Coulomb and exchange pair potentials, and orthonormality
constraints as a coupled nonlinear system on a hierarchy of finite-element
spaces. Forced Gram targets, occupied-frame alignment, and auxiliary-variable
reconstruction provide constraint compatibility and occupied-rotation
equivariance under suitable assumptions. Analysis of the full self-consistent
response identifies an elliptic kinetic principal part, motivating the multilevel treatment. The resulting solver, FAS-H, combines FAS
cycles with a residual-triggered horizontal Newton correction on the finest
grid. Numerical experiments on eight closed-shell systems show convergence in
4--11 cycles across three grid levels. Across 24 system--grid comparisons,
FAS-H is 1.60--5.02 times faster than SCF and uses 68--91\% less weighted FFT 
work, while standalone horizontal Newton is faster in some cases.
\end{abstract}



\section{Introduction}
\label{sec:introduction}

\subsection{Background}

The restricted Hartree--Fock (RHF) method approximates the electronic ground state of a closed-shell system by minimizing an energy functional over orthonormal occupied orbitals. The resulting equations are nonlinear because the Coulomb and nonlocal exchange operators depend on the occupied orbitals. In molecular quantum chemistry, these orbitals are commonly expanded in atom-centered Gaussian basis functions, which provide compact representations and permit analytic evaluation of molecular integrals~\cite{Boys1950,HelgakerJorgensenOlsen2000}. As an alternative, real-space discretizations permit systematic spatial refinement. When combined with multiscale solvers, they can exploit coarse-grid treatment of global constraints and adaptive local refinement, offering the potential for efficient computation~\cite{Beck2000}. Exploiting these advantages for RHF, however, requires addressing orbital orthonormality and nonlocal exchange.

The standard self-consistent-field (SCF) iteration solves the RHF equations by constructing the Fock operator from the current occupied orbitals and solving the resulting linear eigenvalue problem to update them~\cite{Roothaan1951,Pulay1982,KudinScuseriaCances2002}. Direct minimization offers an alternative by optimizing the Hartree--Fock energy under orbital orthonormality constraints~\cite{VanVoorhisHeadGordon2002,HuJiangLinWenYuan2019,Dinvay2026}. For real-space HF, the choice of nonlinear solver is only one part of the algorithm; spatial discretization and efficient evaluation of Coulomb and exchange interactions are also central to solver design. Alizadegan, Hsia, and Martinez developed a divide-and-conquer finite-element HF method~\cite{AlizadeganHsiaMartinez2010}. Braun performed three-dimensional finite-element HF calculations for atoms and small molecules~\cite{Braun2014}. Haubenwallner and Heller used quantics tensor trains for fully numerical HF calculations of atoms and small molecules~\cite{HaubenwallnerHeller2025}. MRChem combines an adaptive multiwavelet representation with an integral formulation based on the Green's function of the shifted kinetic-energy operator~\cite{MRChem2023}. For efficient exchange evaluation, Boffi, Jain, and Natan introduced projection-based compression~\cite{BoffiJainNatan2016}.

Multilevel methods have also been used in real-space HF in several forms. As early as 1992, Davstad combined high-order finite-difference discretization with multigrid and Orthomin iterations to solve diatomic HF equations~\cite{Davstad1992}. Using a finite-element discretization, Beck, Heinemann, and Kolb later applied multigrid with conjugate-gradient smoothing to the linear orbital and Poisson equations within an outer SCF iteration~\cite{BeckHeinemannKolb2003}. Khoromskij, Khoromskaia, and Flad developed multilevel tensor representations for HF calculations within SCF iterations~\cite{KhoromskijKhoromskaiaFlad2011}. Pan and Lindsey combined adaptive discontinuous Galerkin basis sets with multigrid preconditioning for both Poisson equations and the linear eigenvalue problems arising within SCF~\cite{PanLindsey2025}. More recently, Xu, Li and Liang combined adaptive finite-element refinement with multilevel correction, using large-scale linearized boundary-value solves to construct low-dimensional spaces in which the nonlinear HF problem is solved by SCF~\cite{XuLiLiang2026,XuLiangLi2026}.

Related nonlinear multigrid methods, including the full approximation scheme (FAS), have been developed for real-space electronic-structure models with local effective potentials, such as the Schr\"odinger--Poisson and Kohn--Sham equations. Costiner and Ta'asan developed simultaneous multigrid techniques for the Schr\"odinger--Poisson system, coupling orbital, potential, and constraint equations while treating clustered eigenspaces~\cite{CostinerTaasan1995}. For the Kohn--Sham equations, FAS and full multigrid methods combine nonlinear updates with coarse-grid orthogonalization or Ritz projection~\cite{WangBeck2000,WijesekeraFengBeck2003,WijesekeraFengBeck2007}. Localized nonorthogonal formulations further reduce the cost of global constraints~\cite{FengBeck2006}, while multigrid relaxation has also been developed for exact-exchange optimized-effective-potential formulations~\cite{Kummel2004}.

Taken together, these two lines of work demonstrate the potential of multilevel methods for electronic-structure calculations. To our knowledge, however, FAS has not been applied directly to a coupled real-space RHF formulation in which the orbital equations, the Poisson equations for the Coulomb and exchange pair potentials, and the orbital orthonormality constraints are treated as a single nonlinear system.

\subsection{Main idea and contributions}

We formulate the RHF equations as a coupled nonlinear system by introducing
Poisson pair potentials for the Coulomb and exchange interactions, with
orbital orthonormality imposed as part of the
system~\cite{BeckHeinemannKolb2003}. The mutual coupling of the orbital,
pair-potential, and constraint equations motivates a unified nonlinear
multilevel treatment: we investigate FAS applied directly to this system
on nested finite-element spaces~\cite{Brandt1977}, rather than using
multilevel methods solely to accelerate subproblems within an outer SCF
iteration. Since orthogonal rotations of the occupied orbitals leave the
physical state unchanged, we align the coarse- and fine-grid occupied
frames before constructing coarse-grid corrections.

The main contributions of this work are as follows.

\begin{enumerate}
\item We construct a structure-preserving FAS method for the coupled
RHF system. The formulation incorporates forced Gram constraints, 
occupied-frame alignment, and consistent reconstruction of auxiliary variables. 
We establish its constraint consistency and conditional
equivariance under occupied-orbital rotations.
\item We derive the full self-consistent RHF response and establish, 
under suitable field and discretization-stability assumptions, 
a mesh-uniform G\aa rding
estimate identifying its elliptic principal part and lower-order Coulomb
and exchange terms.
\end{enumerate}

Numerical experiments compare FAS-H with SCF and standalone horizontal Newton 
on common VGF discretizations.
Eight closed-shell systems are tested across three grid levels, with 24 
system--grid comparisons of computational cost. 
Additional experiments examine the effects of smoothing and coarse-grid 
correction on different spatial error components.
FAS-H consistently outperforms SCF in the reported comparisons, while 
standalone horizontal Newton is faster in some cases. These results concern 
convergence to common discrete stationary states rather than physical accuracy.

Section~\ref{sec:model} presents the RHF model and its coupled
formulation. Section~\ref{sec:discrete-fas} develops the discrete
coupled FAS method, including its structural properties, solver
realization, and horizontal response correction.
Section~\ref{sec:local-response} examines the linearized RHF response
and its elliptic structure in relation to multigrid.
Section~\ref{sec:numerics} reports grid-refinement experiments,
comparisons with SCF and horizontal Newton, and probes of the
multigrid mechanism. Section~\ref{sec:conclusion} summarizes the
findings and discusses future work.

\section{RHF model}
\label{sec:model}

We consider a closed-shell system of \(2N\) electrons with fixed nuclei and
work in atomic units.
Let \(\Phi=(\phi_1,\ldots,\phi_N)\) denote the real occupied spatial orbitals,
with \(\phi_i\in H^1(\R^3)\), satisfying
\[
\int_{\R^3}\phi_i(x)\phi_j(x)\,dx=\delta_{ij},
\qquad 1\le i,j\le N.
\]
The one-electron operator is
\[
h=-\frac12\Delta+V_{\rm nuc},
\qquad
V_{\rm nuc}(x)=-\sum_{a=1}^{M}\frac{Z_a}{|x-R_a|},
\]
where \(M\) is the number of nuclei, and \(Z_a\) and \(R_a\) denote the
charge and position of nucleus \(a\), respectively.
The RHF electronic energy is
\[
E_{\rm RHF}(\Phi)
=2\sum_{i=1}^N\langle\phi_i,h\phi_i\rangle
 +\sum_{i,j=1}^N\left[2(ii|jj)-(ij|ji)\right].
\]
Here the two-electron integrals are defined by
\[
(ij|kl)
=\iint_{\R^3\times\R^3}
\frac{\phi_i(x)\phi_j(x)\phi_k(y)\phi_l(y)}
{|x-y|}\,dx\,dy .
\]
The RHF ground-state problem is to minimize \(E_{\rm RHF}\) over the real
occupied orbitals subject to the orthonormality constraints.

Stationarity of the constrained energy yields the RHF equations
\[
\left[h+\sum_{j=1}^N(2J_j-K_j)\right]\phi_i
=\sum_{j=1}^N\phi_j\Lambda_{ji},
\qquad i=1,\ldots,N.
\]
Here \(J_j\) and \(K_j\) are the Coulomb and exchange operators defined by
\[
(J_j u)(x)
=\left(\int_{\R^3}\frac{\phi_j(y)^2}{|x-y|}\,dy\right)u(x),
\qquad
(K_j u)(x)
=\phi_j(x)\int_{\R^3}\frac{\phi_j(y)u(y)}{|x-y|}\,dy,
\]
and \(\Lambda\in\R_{\mathrm{sym}}^{N\times N}\) is the matrix of
Lagrange multipliers enforcing orbital orthonormality.
Since the RHF energy is invariant under orthogonal rotations of the occupied
orbitals, we retain the full symmetric multiplier matrix rather than select
canonical orbitals by diagonalizing it.

To express the Coulomb and exchange terms through auxiliary fields, define
the symmetric pair potentials
\[
W_{ij}(x)=\int_{\R^3}
\frac{\phi_i(y)\phi_j(y)}{|x-y|}\,dy,
\qquad W_{ij}=W_{ji}.
\]
These potentials satisfy
\[
J_j\phi_i=W_{jj}\phi_i,
\qquad
K_j\phi_i=W_{ij}\phi_j,
\qquad
-\Delta W_{ij}=4\pi\phi_i\phi_j
\quad\text{in }\R^3.
\]
With the pair potentials defined by the Coulomb integrals above, the RHF
equations are equivalently written as the coupled orbital, Poisson, and
orthonormality system
\begin{equation}
\label{eq:continuous-rhf}
\boxed{
\begin{aligned}
-\frac12\Delta\phi_i+V_{\rm nuc}\phi_i
 +2\left(\sum_{j=1}^N W_{jj}\right)\phi_i
 -\sum_{j=1}^N W_{ij}\phi_j
 &=\sum_{j=1}^N\phi_j\Lambda_{ji},
 &&i=1,\ldots,N,\\
-\Delta W_{ij}&=4\pi\phi_i\phi_j,
 &&1\leq i\leq j\leq N,\\
\int_{\R^3}\phi_i\phi_j&=\delta_{ij},
 &&1\leq i\leq j\leq N.
\end{aligned}}
\end{equation}

In~\eqref{eq:continuous-rhf}, the orbitals determine the source terms
\(4\pi\phi_i\phi_j\) in the pair-potential equations, while the resulting
Coulomb and exchange potentials enter the orbital equations and affect
the orbital updates. The orthonormality constraints additionally couple
all occupied orbitals through global inner products. These dependencies
make the orbital, pair-potential, and constraint equations components of
a single nonlinear RHF problem, whether solved by simultaneous updates
or by block iterations such as SCF.

\section{A coupled FAS method for RHF}
\label{sec:discrete-fas}

\subsection{Discrete RHF model}

For numerical computation, we approximate the whole-space problem on a
bounded domain \(\Omega\subset\R^3\).
The orbitals satisfy homogeneous Dirichlet boundary conditions on
\(\partial\Omega\) and are extended by zero outside \(\Omega\).
The pair potentials remain the free-space Coulomb potentials generated
by the truncated orbital products, so their boundary values are generally
nonzero and are determined by the same Coulomb integrals.

Let
\begin{equation*}
V_0\subset V_1\subset\cdots\subset V_L\subset H_0^1(\Omega)
\end{equation*}
be nested conforming finite-element spaces for the orbitals.
We represent pair sources and potentials in a separate finite-element
space \(Q_\ell\subset L^2(\Omega)\), without homogeneous Dirichlet
boundary conditions. Let \(\Pi_\ell\) specify the representation of orbital
products in \(Q_\ell\), and let \(\mathcal C_\ell\) map a represented source
to the \(L^2\) projection of its free-space Coulomb potential onto \(Q_\ell\).

The discrete RHF problem is to find \(\phi_{\ell,i}\in V_\ell\),
\(W_{\ell,ij}=W_{\ell,ji}\in Q_\ell\), and a symmetric matrix
\(\Lambda_\ell\) such that, for every \(v\in V_\ell\),
\begin{equation}
\label{eq:discrete-weak-rhf}
\begin{aligned}
\frac12(\nabla v,\nabla\phi_{\ell,i})
 +(v,V_{\rm nuc}\phi_{\ell,i})
 &+2\sum_j(\Pi_\ell(v\phi_{\ell,i}),W_{\ell,jj})_{Q,\ell}\\
 &-\sum_j(\Pi_\ell(v\phi_{\ell,j}),W_{\ell,ij})_{Q,\ell}
 =\sum_j(v,\phi_{\ell,j})_{V,\ell}\Lambda_{\ell,ji},\\
W_{\ell,ij}&=\mathcal C_\ell\Pi_\ell(\phi_{\ell,i}\phi_{\ell,j}),\\
(\phi_{\ell,i},\phi_{\ell,j})_{V,\ell}&=\delta_{ij}.
\end{aligned}
\end{equation}
Here the spatial pairings are the \(L^2\) inner products on \(\Omega\),
with their chosen discrete realizations on \(V_\ell\) and \(Q_\ell\).
When \(\Pi_\ell\) is the \(L^2\) projection, the interaction terms reduce
to the usual Galerkin potential terms. The three equations couple the
orbitals, pair potentials, and orthonormality constraints.

\subsection{FAS framework}
\label{sec:fas}

\subsubsection{Complete discrete operator}

We collect the discrete unknowns into
\[
X_\ell=(\Phi_\ell,W_\ell,\Lambda_\ell),
\]
where \(\Phi_\ell\) contains the occupied orbitals, \(W_\ell\) is the
symmetric family of pair potentials, and \(\Lambda_\ell\) is the
symmetric multiplier matrix.

Let \(M_\ell\) and \(K_\ell^{\rm stiff}\) be the orbital mass and
stiffness matrices, and set
\[
A_\ell=M_\ell^{-1}K_\ell^{\rm stiff},
\qquad
G_\ell(\Phi)=\Phi^{\mathsf T}M_\ell\Phi.
\]
We denote the discrete nuclear-potential operator by
\(V_{{\rm nuc},\ell}\).
The orbital and pair-space mass matrices define the inner products
\[
(u,v)_{V,\ell}=u^{\mathsf T}M_\ell v,
\qquad
(s,t)_{Q,\ell}=s^{\mathsf T}M_{Q,\ell}t.
\]

For the interaction terms, define the represented pair sources
\begin{equation*}
s_{\ell,ij}(\Phi)=\Pi_\ell(\phi_{\ell,i}\phi_{\ell,j})
\end{equation*}
and the discrete potential multiplication operator
\(\mathcal M_{\ell,w}\) through
\[
(v,\mathcal M_{\ell,w}u)_{V,\ell}
=(\Pi_\ell(uv),w)_{Q,\ell},
\qquad u,v\in V_\ell,\quad w\in Q_\ell.
\]

The discrete Fock action is
\[
\cH_{\ell,i}(\Phi,W)
=\frac12A_\ell\phi_i
 +V_{{\rm nuc},\ell}\phi_i
 +2\sum_{j=1}^N\mathcal M_{\ell,W_{jj}}\phi_i
 -\sum_{j=1}^N\mathcal M_{\ell,W_{ij}}\phi_j .
\]
With a fixed invertible scaling
\(\mathcal D_\ell:Q_\ell\to Q_\ell\), define
\(F_\ell=(F_{\Phi,\ell},F_{W,\ell},F_{G,\ell})\) by
\begin{equation}
\label{eq:kkt-blocks}
\begin{aligned}
F_{\Phi,\ell,i}(X)
&=\cH_{\ell,i}(\Phi,W)-\sum_j\phi_j\Lambda_{ji},\\
F_{W,\ell,ij}(X)
&=\mathcal D_\ell
  [W_{ij}-\mathcal C_\ell s_{\ell,ij}(\Phi)],\\
F_{G,\ell}(X)&=G_\ell(\Phi)-I.
\end{aligned}
\end{equation}
The scaling \(\mathcal D_\ell\) preserves the pair relation while
fixing its residual representation.
The finest-level problem is
\begin{equation*}
F_L(X_L)=0.
\end{equation*}
We apply FAS to the coupled system, treating the orbitals, pair
potentials, and orthonormality constraints together across the grid
hierarchy. The coarse-grid equations incorporate the restricted
fine-grid residual through the FAS correction defined below.

\subsubsection{FAS coarse-grid equations}

FAS solves the full nonlinear system on each level, modifying the coarse
right-hand side so that the restriction of an exact fine-level solution
already satisfies the coarse equation.
For adjacent levels, let \(I_h^H,I_{Q,h}^H\) restrict orbital and
pair states, and let \(P_H^h,P_{Q,H}^h\) be their prolongations.
The complete state and residual restrictions are
\[
\begin{aligned}
\cI_h^H(\Phi,W,\Lambda)
&=\bigl(I_h^H\Phi,I_{Q,h}^HW,\Lambda\bigr),\\
\cR_h^H(r_\Phi,r_W,r_G)
&=\bigl(R_h^Hr_\Phi,R_{Q,h}^Hr_W,r_G\bigr).
\end{aligned}
\]
The residual restrictions are the mass adjoints of prolongation:
\[
R_h^H=M_H^{-1}(P_H^h)^{\mathsf T}M_h,
\qquad
R_{Q,h}^H=M_{Q,H}^{-1}(P_{Q,H}^h)^{\mathsf T}M_{Q,h}.
\]

For a current fine-level state \(\overline X_h\), define
\(Y_H=\cI_h^H\overline X_h\) and the tau correction
\begin{equation}
\label{eq:tau-form}
\tau_H=F_H(Y_H)-\cR_h^HF_h(\overline X_h).
\end{equation}
The coarse equation is \(F_H(X_H)=B_H\), with
\begin{equation}
\label{eq:fas-rhs}
B_H=\cR_h^HB_h+\tau_H
   =F_H(Y_H)+\cR_h^H[B_h-F_h(\overline X_h)].
\end{equation}
Thus, if \(F_h(\overline X_h)=B_h\), then \(F_H(Y_H)=B_H\).
The tau correction accounts for the difference between the coarse
operator and the restricted fine operator; it equals \(B_H\) only when
\(B_h=0\).

We therefore use the forced coupled system on every level:
\begin{equation}
\label{eq:forced-level}
F_\ell(X_\ell)=B_\ell,\qquad
B_\ell=(B_{\Phi,\ell},B_{W,\ell},B_{G,\ell}),
\end{equation}
with \(B_L=0\) on the finest level.  In block form,
\eqref{eq:forced-level} is
\begin{equation}
\label{eq:forced-blocks}
\begin{aligned}
\cH_{\ell,i}(\Phi,W)-\sum_j\phi_j\Lambda_{ji}
    &=B_{\Phi,\ell,i},\\
\mathcal D_\ell\bigl[W_{ij}-\mathcal C_\ell s_{\ell,ij}(\Phi)\bigr]
    &=B_{W,\ell,ij},\\
G_\ell(\Phi)&=S_\ell,\qquad S_\ell:=I+B_{G,\ell}.
\end{aligned}
\end{equation}
The constraint forcing \(B_{G,\ell}\) is symmetric, and we require
the prescribed Gram matrix \(S_\ell=I+B_{G,\ell}\) to be positive definite.
Thus coarse-level orbital updates must preserve \(S_\ell\), which
generally differs from \(I\).

\subsubsection{Orbital alignment}
\label{sec:gauge-correction}

Occupied orbitals are determined only up to an orthogonal rotation,
so coarse- and fine-grid frames can represent the same occupied subspace
using different bases. Before forming the correction, we align the
coarse solution with the restricted fine-grid frame in their common
coarse space. This makes their orbital representations comparable and
prevents an arbitrary basis rotation from being interpreted as physical
subspace error. This representation issue is distinct from the
performance comparison between coupled FAS and SCF.

For \(Q\in O(N)\), the corresponding transformation of the coupled state is
\[
\Phi\mapsto\Phi Q,\qquad
W\mapsto Q^{\mathsf T}WQ,\qquad
\Lambda\mapsto Q^{\mathsf T}\Lambda Q.
\]
This transformation preserves the electron density, RHF energy,
and occupied subspace.
The residual blocks transform as
\begin{equation}
\label{eq:gauge-residual}
F_\Phi\mapsto F_\Phi Q,\qquad
F_W\mapsto Q^{\mathsf T}F_WQ,\qquad
F_G\mapsto Q^{\mathsf T}F_GQ.
\end{equation}
For a forced system, the same transformation must be applied to
\(B_\Phi,B_W,B_G\), with \(S\mapsto Q^{\mathsf T}SQ\).
Thus covariance relates jointly rotated states and right-hand sides,
rather than states with fixed forcing.

Let \(Y_H=\cI_h^H\overline X_h\) and let \(Z_H\) approximately solve
\(F_H(Z_H)=B_H\), starting from \(Y_H\).
To compare their occupied subspaces independently of their Gram matrices,
form the orthonormal frames
\[
\widehat Y_H=Y_{\Phi,H}G_H(Y_{\Phi,H})^{-1/2},
\qquad
\widehat Z_H=Z_{\Phi,H}G_H(Z_{\Phi,H})^{-1/2}.
\]
We align the coarse solution with the reference frame by solving
the mass-weighted orthogonal Procrustes problem
\begin{equation}
\label{eq:procrustes-problem}
Q_\star
=\arg\min_{Q^{\mathsf T}Q=I}
\|\widehat Z_HQ-\widehat Y_H\|_{M_H,F}.
\end{equation}
If the overlap matrix is nonsingular, the minimizer is unique
and is obtained from its singular value decomposition:
\[
\widehat Z_H^{\mathsf T}M_H\widehat Y_H
=U\Sigma V^{\mathsf T},
\qquad
Q_\star=UV^{\mathsf T}.
\]
Apply this rotation to the complete coarse state:
\[
Z_H^{\rm al}
=\left(
Z_{\Phi,H}Q_\star,\,
Q_\star^{\mathsf T}Z_{W,H}Q_\star,\,
Q_\star^{\mathsf T}Z_{\Lambda,H}Q_\star
\right),
\]
and transform the coarse forcing by the same rules.
Write \(\widehat Z_H^{\rm al}=\widehat Z_HQ_\star\).
These rotations change only the coarse coordinates; the fine-level
forcing remains unchanged.

After Procrustes alignment, we use the aligned component correction,
which prolongs the full aligned coarse-frame difference:
\begin{equation}
\label{eq:component-orbital-correction}
\Phi_h^+
=\cR_{h,S_h}\left[
\overline\Phi_h+\omega_cP_H^h
(Z_{\Phi,H}^{\rm al}-Y_{\Phi,H})
\right],
\end{equation}
where \(S_h=I+B_{G,h}\), \(0<\omega_c\leq1\), and
\(\cR_{h,S_h}\) is the prescribed-Gram retraction
\begin{equation}
\label{eq:gram-retraction}
\cR_{\ell,S}(\Psi)
=\Psi G^{-1/2}
\left(G^{1/2}SG^{1/2}\right)^{1/2}G^{-1/2},
\qquad G=\Psi^{\mathsf T}M_\ell\Psi.
\end{equation}
For full-rank \(\Psi\) and \(S\succ0\), the symmetric positive-definite
square roots are well-defined, and
\(G_\ell(\cR_{\ell,S}(\Psi))=S\).

\begin{remark}[Projected alternative]
A projected correction can instead be constructed from the normalized
aligned frame difference:
\begin{equation}
\label{eq:coarse-horizontal-chord}
\eta_H=
\left(I-\widehat Y_H\widehat Y_H^{\mathsf T}M_H\right)
\left(\widehat Z_H^{\rm al}-\widehat Y_H\right).
\end{equation}
After prolongation, a second horizontal projection gives
\[
\widehat\Phi_h=\overline\Phi_hS_h^{-1/2},\qquad
\widetilde\eta_h=P_H^h\eta_H,
\]
\begin{equation*}
\widehat\eta_h=
\left(I-\widehat\Phi_h
\widehat\Phi_h^{\mathsf T}M_h\right)
\widetilde\eta_h,\qquad
\delta\Phi_h=\widehat\eta_hS_h^{1/2}.
\end{equation*}
The lift by \(S_h^{1/2}\) gives
\[
\overline\Phi_h^{\mathsf T}M_h\delta\Phi_h
 +\delta\Phi_h^{\mathsf T}M_h\overline\Phi_h=0.
\]
The corresponding update is
\begin{equation}
\label{eq:geometric-orbital-correction}
\Phi_h^+=\cR_{h,S_h}
\left(\overline\Phi_h+\omega_c\delta\Phi_h\right).
\end{equation}
It also restores the prescribed Gram matrix. The numerical experiments
in this paper use the component correction~\eqref{eq:component-orbital-correction}.
\end{remark}

Both directions are covariant under independent changes of coarse-frame
representatives. The normalized aligned chord and its horizontal projection
differ quadratically, and the fine retraction preserves a horizontal
increment to first order; see Proposition~\ref{prop:fas-structure}.

We require the spatial transfers to commute with constant occupied-orbital
rotations, so that the FAS right-hand side transforms with the state.

\begin{proposition}[Constraint compatibility and aligned corrections]
\label{prop:fas-structure}
Assume that the orbital mass matrices are symmetric positive definite.
\begin{enumerate}
\item If $G_h(\overline\Phi_h)=S_h$ and the Gram residual is injected,
then $S_H=G_H(I_h^H\overline\Phi_h)$. This target is positive definite
exactly when the restricted frame has full column rank.
\item Assume that the coarse frames and the fine correction trial frame have
full column rank, $S_h$ is positive definite, the normalized coarse overlap
is nonsingular, and the spatial transfers commute with orbital rotations.
Under $Y\mapsto YQ_Y$, $Z\mapsto ZQ_Z$,
$\overline\Phi_h\mapsto\overline\Phi_hQ_Y$, and
$S_h\mapsto Q_Y^{\mathsf T}S_hQ_Y$, either aligned correction satisfies
$\Phi_h^+\mapsto\Phi_h^+Q_Y$.
\item Assume that the coarse frames have full column rank and their
normalized overlap is nonsingular. With
$\widehat Z_{\rm al}=\widehat ZQ_\star$,
$C=\widehat Y^{\mathsf T}M_H\widehat Z_{\rm al}$, and $\eta$ as
in~\eqref{eq:coarse-horizontal-chord},
\begin{align*}
 C&=V\Sigma V^{\mathsf T}>0,\qquad
 \widehat Y^{\mathsf T}M_H\eta=0,\\
 \widehat Z_{\rm al}-\widehat Y
 &=\eta+\widehat Y(C-I),\qquad
 C=(I-\eta^{\mathsf T}M_H\eta)^{1/2}.\end{align*}
Thus $C-I=O(\|\eta\|_{M_H,F}^2)$ near coincident subspaces. If
$S_h$ is positive definite and $G_h(\overline\Phi_h)=S_h$, the fine
horizontal lift satisfies
\begin{equation*}
\cR_{h,S_h}(\overline\Phi_h+t\delta\Phi_h)
=\overline\Phi_h+t\delta\Phi_h+O(t^2).
\end{equation*}
\end{enumerate}
\end{proposition}
\begin{proof}
The Gram block of~\eqref{eq:fas-rhs} gives
$S_H=G_H(I_h^H\overline\Phi_h)+B_{G,h}-[G_h(\overline\Phi_h)-I]$;
the last two terms cancel, and SPD mass makes the resulting Gram matrix
positive definite exactly at full rank.

Orthogonal covariance of SPD roots gives
$\widehat{YQ_Y}=\widehat YQ_Y$ and
$\cR_{h,Q_Y^{\mathsf T}S_hQ_Y}(\Psi Q_Y)=\cR_{h,S_h}(\Psi)Q_Y$.
Uniqueness of the polar factor gives
$Q_\star'=Q_Z^{\mathsf T}Q_\star Q_Y$, so both the aligned component
difference and horizontal lift transform by right multiplication by $Q_Y$.
This proves covariance of either retracted correction.

The SVD gives $C=V\Sigma V^{\mathsf T}>0$ and
$\widehat Z_{\rm al}=\widehat YC+\eta$ with horizontal $\eta$.
Taking its Gram yields $I=C^2+\eta^{\mathsf T}M_H\eta$.
Smoothness of the SPD root proves the quadratic estimate. Likewise,
$G_h(\overline\Phi_h+t\delta\Phi_h)
=S_h+t^2\delta\Phi_h^{\mathsf T}M_h\delta\Phi_h$;
the retraction factor is $I+O(t^2)$.

\end{proof}
\subsection{Complete FAS algorithm}
\label{sec:complete-fas}

After either orbital correction, we reconstruct the pair potentials
using the unchanged fine-level forcing:
\begin{equation}
\label{eq:pair-refresh}
W_{ij}^+=\mathcal C_hs_{h,ij}(\Phi_h^+)
+\mathcal D_h^{-1}B_{W,h,ij}.
\end{equation}
Retaining the forcing term ensures that the updated fields satisfy
the active pair equations.

The symmetric multiplier is then chosen to minimize the orbital residual:
\[
\Lambda_h^+
=\arg\min_{\Lambda=\Lambda^{\mathsf T}}
\|\cH_h(\Phi_h^+,W_h^+)-B_{\Phi,h}-\Phi_h^+\Lambda\|_{M_h,F}^2.
\]
Equivalently, it is the unique symmetric solution of
\begin{equation}
\label{eq:sylvester-update}
S_h\Lambda_h^++\Lambda_h^+S_h
=2\sym\!\left[
(\Phi_h^+)^{\mathsf T}M_h
\bigl(\cH_h(\Phi_h^+,W_h^+)-B_{\Phi,h}\bigr)
\right].
\end{equation}
Uniqueness follows from \(S_h\succ0\).

Thus the orbital coarse correction also updates the pair potentials and
multiplier through reconstruction, while restoring the active Gram
constraint. All three blocks remain tied to the same forced nonlinear
problem.

Let \(\mathcal S_\ell^\nu(X;B)\) denote \(\nu\) nonlinear smoothing
sweeps for \(F_\ell(X)=B\).
The resulting FAS V-cycle for the coupled RHF system is summarized in
Algorithm~\ref{alg:fas-cycle}.

\begin{algorithm}
\caption{Structure-preserving FAS V-cycle for RHF}
\label{alg:fas-cycle}
\begin{algorithmic}[1]
\Require level $\ell$, state $X_\ell$, forcing $B_\ell$, sweeps $\nu_1,\nu_2$
\If{$\ell=0$}
 \State \Return $\operatorname{CoarseSolve}(X_0;B_0)$
\EndIf
\State $H\gets\ell-1$
\State $\overline X\gets\mathcal S_\ell^{\nu_1}(X_\ell;B_\ell)$
\State $Y\gets\cI_\ell^H\overline X$
\State $B_H\gets F_H(Y)+\cR_\ell^H[B_\ell-F_\ell(\overline X)]$
\State $Z\gets\operatorname{FASCycle}(H,Y,B_H)$
\State align $Z$ and its coarse forcing with the reference $Y$
\State $d\gets P_H^\ell(Z_{\Phi}^{\rm al}-Y_{\Phi})$ as in~\eqref{eq:component-orbital-correction}
\State choose an admissible correction weight $\omega_c$
\State $\Phi^+\gets\cR_{\ell,S_\ell}(\overline\Phi+\omega_c d)$
\State reconstruct $W^+$ by~\eqref{eq:pair-refresh} with the original $B_{W,\ell}$
\State update $\Lambda^+$ by~\eqref{eq:sylvester-update} with the original $B_{\Phi,\ell}$
\State \Return $\mathcal S_\ell^{\nu_2}((\Phi^+,W^+,\Lambda^+);B_\ell)$
\end{algorithmic}
\end{algorithm}

We monitor convergence using the coupled residual norm
\begin{equation}
\label{eq:product-norm}
\|R\|_{\mathcal Y_\ell}^2
=\|R_\Phi\|_{M_\ell,F}^2
 +\sum_i\|R_{W,ii}\|_{M_{Q,\ell}}^2
 +2\sum_{i<j}\|R_{W,ij}\|_{M_{Q,\ell}}^2
 +\|R_G\|_F^2,
\end{equation}
where \(R=F_\ell(X_\ell)-B_\ell\).
The factor two accounts for the symmetric off-diagonal pair fields.

On the physical finest level, we use the residual-triggered correction
defined in Section~\ref{sec:physical-correction} after each complete
cycle with \(B_L=0\), until
\(\|F_L(X_L)\|_{\mathcal Y_L}\leq\varepsilon\).
Experimental parameters are given in Section~\ref{sec:implementation}.

The following proposition establishes conditions under which the
complete FAS cycle inherits rotation equivariance and preserves
exact solutions of the forced system.

\begin{proposition}[Equivariance and fixed-point preservation]
\label{prop:fas-cycle-structure}
Let \(T_Q\) denote the joint rotation of the state and forcing.
Assume that the coupled operators are rotation-covariant and the spatial
transfers commute with rotations. Assume also that the corrections are
well-defined under the conditions of Proposition~\ref{prop:fas-structure}.
Then the residual norm~\eqref{eq:product-norm} is rotation-invariant,
and the following hold.
\begin{enumerate}
\item If smoothing and the coarsest solver commute with joint rotations,
and all adaptive parameter choices and admissibility decisions are
rotation-invariant, then
\[
\operatorname{FASCycle}(T_QX;T_QB)
=T_Q\operatorname{FASCycle}(X;B),
\]
\item If smoothing and the coarsest solver preserve exact forced solutions,
then every admissible solution of \(F_\ell(X_\ell)=B_\ell\) is a
fixed point of the complete cycle.
\end{enumerate}
\end{proposition}
\begin{proof}
Cyclicity of the trace gives invariance of the residual norm.
The transfers, aligned correction, pair reconstruction, and unique
symmetric multiplier update commute with joint rotations.
Together with the assumptions on smoothing, the coarsest solver,
and adaptive decisions, induction over levels proves cycle equivariance.

If \(F_h(X_h)=B_h\), smoothing leaves \(X_h\) unchanged and
\eqref{eq:fas-rhs} gives \(B_H=F_H(Y_H)\).
Induction from the coarsest level leaves \(Y_H\) unchanged, so both
orbital correction strategies produce zero increments.
Retraction, pair reconstruction, and the unique symmetric multiplier
update then preserve \(X_h\), as does post-smoothing.
\end{proof}

\subsection{Solver realization}
\label{sec:solver-realization}

We realize the FAS cycle using orbital block Gauss--Seidel relaxation on
each noncoarsest level and a reduced Newton--Krylov solve on the coarsest
level. The relaxation updates the orbitals while maintaining the active
constraints and pair equations; the coarse solver eliminates the pair
potentials and solves the remaining coupled orbital and Gram equations.

Each Gauss--Seidel candidate starts with an update of the orbital equations,
followed by restoration of the coupled variables:
\begin{enumerate}
\item With \(W\) and \(\Lambda\) fixed, update the occupied orbital
values together at each spatial vertex.
\item Retract the updated orbital frame using~\eqref{eq:gram-retraction}
to satisfy the active Gram target \(G_\ell(\Phi)=S_\ell\).
\item Reconstruct the pair potentials using~\eqref{eq:pair-refresh}
with the active forcing \(B_{W,\ell}\).
\item Update the symmetric multiplier using~\eqref{eq:sylvester-update}
with the active forcing \(B_{\Phi,\ell}\).
\end{enumerate}

On the coarsest level, the active pair equations determine the pair
potentials from the orbitals and the pair forcing. We eliminate these
potentials using~\eqref{eq:pair-refresh}, leaving a coupled system for the
orbitals and the symmetric multiplier. Define
\[
\begin{aligned}
W_B(\Phi)&=\mathcal C_\ell s_\ell(\Phi)+\mathcal D_\ell^{-1}B_W,\\
r_\Phi&=\cH_\ell(\Phi,W_B(\Phi))-\Phi\Lambda-B_\Phi,\qquad
r_G=G_\ell(\Phi)-S_\ell.
\end{aligned}
\]
The Newton correction \((E,\Delta\Lambda)\) satisfies
\begin{equation}
\label{eq:forced-newton-system}
\begin{aligned}
\mathcal J_\ell E-\Phi\Delta\Lambda&=-r_\Phi,\qquad
                                  \Delta\Lambda=\Delta\Lambda^{\mathsf T},\\
\Phi^{\mathsf T}M_\ell E+E^{\mathsf T}M_\ell\Phi&=-r_G,
\end{aligned}
\end{equation}
Here \(\mathcal J_\ell\) is the derivative of
\(\cH_\ell(\Phi,W_B(\Phi))-\Phi\Lambda\) with respect to \(\Phi\), with
\(\Lambda\) and the forcing held fixed. Its matrix-free action includes
the induced response of the pair potentials. We solve the coupled Newton
system by GMRES, using a preconditioner based on shifted kinetic operators
and a small Sylvester Schur complement. The linear residual is recomputed
after each solve.

\subsection{Horizontal response correction}
\label{sec:physical-correction}

In addition to the coarse-level Newton solve, we apply a horizontal
Newton correction on the finest grid when a FAS cycle does not
sufficiently reduce the physical residual. On this grid the
forcing is zero, the Gram target is \(I\), and the pair potentials are
determined by the orbitals. We call the resulting solver FAS-H.

Orthogonal rotations of the occupied orbitals represent the same physical
state. We therefore compute the Newton correction in the horizontal space,
excluding directions within the occupied subspace. For a mass-orthonormal
frame \(\Phi\), define
\[
Q_\Phi=I-\Phi\Phi^{\mathsf T}M_L,\qquad
W=\mathcal C_Ls_L(\Phi),\qquad
\Lambda=\sym\!\left[\Phi^{\mathsf T}M_L\cH_L(\Phi,W)\right].
\]
The horizontal Newton direction approximately satisfies
\begin{equation}
\label{eq:physical-horizontal-newton}
Q_\Phi\mathcal J_\Phi E=-Q_\Phi r_\Phi,
\qquad \Phi^{\mathsf T}M_LE=0,
\end{equation}
Here \(\mathcal J_\Phi\) is the orbital response with the current multiplier
held fixed, including the induced response of the pair potentials. The
multiplier derivative lies in the occupied subspace and is removed by
\(Q_\Phi\). We solve the projected equation using GMRES in mass-whitened
coordinates with a projected kinetic preconditioner. The horizontal
projection applies to the unforced physical problem; the forced coarse
problems retain the joint orbital--multiplier system of
\eqref{eq:forced-newton-system}.

After each call to the finest-grid FAS cycle returns, we compare the complete
physical residual with its value at the beginning of the cycle. If the residual
remains above the prescribed tolerance and is at least a fraction \(\theta\)
of its entering value, FAS-H attempts at most one horizontal Newton step. For
each trial, the orbital frame is retracted and the pair potentials and
multiplier are rebuilt; the first valid trial that decreases the complete
physical residual is accepted. Algorithm~\ref{alg:fas-horizontal} summarizes
the resulting outer iteration.

\begin{algorithm}
\caption{One outer iteration of the physical FAS-H solver}
\label{alg:fas-horizontal}
\begin{algorithmic}[1]
\Require admissible physical state $X$, finest level $L$, tolerance $\varepsilon$, trigger $\theta$
\State $r_{\rm in}\gets\|F_L(X)\|_{\mathcal Y_L}$
\State $Y\gets\operatorname{FASCycle}(L,X,0)$ using Algorithm~\ref{alg:fas-cycle}
\State $r_Y\gets\|F_L(Y)\|_{\mathcal Y_L}$
\If{$r_Y>\varepsilon$ and $r_Y\ge\theta r_{\rm in}$}
 \State compute at most one bounded, backtracked horizontal Newton step from $Y$, obtaining $Z$
 \If{$\|F_L(Z)\|_{\mathcal Y_L}<r_Y$}
  \State $Y\gets Z$
 \EndIf
\EndIf
\State \Return $Y$
\end{algorithmic}
\end{algorithm}

Retraction and reconstruction restore the physical Gram and pair equations
for each accepted correction. Under covariant linear solves and
rotation-invariant adaptive decisions, the horizontal correction also
respects occupied-orbital rotations.

\section{RHF response and elliptic structure}
\label{sec:local-response}

To understand the multigrid treatment of the coupled RHF system,
we examine its linearized response near a discrete stationary state.
Under mesh-independent bounds on the stationary solutions and suitable
stability assumptions on the discrete operators, this response retains
the elliptic kinetic principal part, while Coulomb and exchange
contributions are lower-order terms.
This structure motivates the complementary roles of smoothing
and coupled coarse-grid correction.

\subsection{Linearized response of the occupied subspace}
\label{sec:schur-response}

We first examine how changes in the orbitals and their induced potentials
affect the RHF equations, then remove the freedom associated with
occupied-orbital rotations.

Let \(X_*=(\Phi_*,W_*,\Lambda_*)\) be a discrete stationary state
satisfying \(F_h(X_*)=0\). An orbital perturbation
\(E=(e_1,\ldots,e_N)\) induces the pair-potential response
\begin{equation}
\label{eq:pair-response}
\dot W_{ij}[E]
=\mathcal C_h\Pi_h(e_i\phi_{*,j}+\phi_{*,i}e_j),
\end{equation}
with the discrete maps \(\Pi_h\) and \(\mathcal C_h\) held fixed.
Substituting this potential response into the orbital equations gives
the full self-consistent response
\begin{equation}
\label{eq:rhf-response}
(J_*E)_i=(\mathcal A_*E)_i
+2\sum_j\mathcal M_{h,\dot W_{jj}[E]}\phi_{*,i}
-\sum_j\mathcal M_{h,\dot W_{ij}[E]}\phi_{*,j},
\end{equation}
where \(\mathcal A_*E=\cH_h(E,W_*)-E\Lambda_*\) is the fixed-field
response. The remaining terms describe the induced Coulomb and exchange
feedback.

To isolate changes in the occupied subspace, we restrict \(E\) to the
horizontal space \(\mathcal T_h=\{E:\Phi_*^{\mathsf T}M_hE=0\}\).
These perturbations preserve orthonormality to first order and exclude
infinitesimal occupied rotations. The linearized orbital residual also
contains a multiplier perturbation, \(J_*E-\Phi_*\dot\Lambda\).
Projecting onto the horizontal space removes this multiplier term and gives
\begin{equation}
\label{eq:horizontal-response}
L_h=P_*J_*\big|_{\mathcal T_h},
\qquad P_*=I-\Phi_*\Phi_*^{\mathsf T}M_h.
\end{equation}
Thus \(L_h\) describes the self-consistent response of the occupied subspace
independently of the multiplier perturbation.

The occupied-subspace response also describes the local curvature
of the constrained RHF energy.
If \(\mathcal C_h\) is self-adjoint in the pair-space inner product
and the energy and orbital equations use consistent weak product
pairings, then \(L_h\) is mass-self-adjoint on \(\mathcal T_h\), with
\begin{equation}
\label{eq:constrained-hessian}
D^2\mathcal L(\Phi_*)[E,E]=4(E,L_hE)_{M_h,F},
\qquad E\in\mathcal T_h.
\end{equation}
Here
\[
\mathcal L(\Phi)
=E_h(\Phi)-2\operatorname{tr}
[\Lambda_*(G_h(\Phi)-I)]
\]
is the discrete RHF Lagrangian, with \(E_h\) the electronic energy
and \(\Lambda_*\) held fixed.
The derivation is given in Appendix~\ref{app:constrained-curvature}.

\subsection{Elliptic structure and spatial scales}
\label{sec:elliptic-response}

To identify the spatial principal part, we separate the horizontal
response into kinetic, nuclear, fixed-field, and induced-potential terms:
\begin{equation}
\label{eq:response-elliptic-split}
L_hE=P_*\left[
\tfrac12A_hE+V_{{\rm nuc},h}E+\mathcal V_hE+\mathcal K_hE
\right].
\end{equation}
Here \(\mathcal V_h=\mathcal A_*-\tfrac12A_h-V_{{\rm nuc},h}\)
contains the fixed pair-potential terms and multiplier shift,
while \(\mathcal K_h=J_*-\mathcal A_*\) contains the induced Coulomb
and exchange response.

\begin{proposition}[Elliptic principal part of the RHF response]
\label{prop:elliptic-response}
Consider a family of discrete RHF stationary states on a fixed domain
\(\Omega\), with fixed \(N\), and let \(L_h\) be defined
by~\eqref{eq:horizontal-response}.
Write
\[
\|E\|_h=\|E\|_{M_h,F},
\qquad
|E|_{1,h}^2=\sum_i(e_i,A_he_i)_{V,h}.
\]
Assume that the discrete mass and stiffness forms are uniformly
equivalent to the conforming \(L^2\) and gradient forms.
Assume that \(\|\Lambda_*\|_2\) is bounded independently of \(h\), and that
the Coulomb maps \(\mathcal C_h\), represented product maps
\(v\mapsto\Pi_h(\phi_{*,i}v)\), and weak multiplication operators
\(\mathcal M_{h,W_{*,ij}}\) have \(h\)-independent bounds in their
respective mass norms.

For the nuclear potential, assume that for every \(\epsilon>0\),
\[
\left|(E,V_{{\rm nuc},h}E)_{M_h,F}\right|
\leq \epsilon|E|_{1,h}^2+c_\epsilon\|E\|_h^2,
\]
with \(c_\epsilon\) independent of \(h\).
Then, for every \(\epsilon>0\), there exists \(C_\epsilon\)
independent of \(h\) such that
\begin{equation}
\label{eq:response-garding}
\left|(E,L_hE)_{M_h,F}-\tfrac12|E|_{1,h}^2\right|
\le\epsilon|E|_{1,h}^2+C_\epsilon\|E\|_h^2,
\end{equation}
for all \(E\in\mathcal T_h\).
\end{proposition}
\begin{proof}
The operator and multiplier bounds imply
\[
\|\mathcal V_hE\|_h\leq C_V\|E\|_h,
\qquad
\|\mathcal K_hE\|_h\leq C_K\|E\|_h,
\]
with \(C_V,C_K\) independent of \(h\).
For \(\mathcal K_h\), this follows by composing the bounded product
maps, their mass adjoints, and the Coulomb map.
For \(E\in\mathcal T_h\), orthogonality of \(P_*\) gives
\((E,P_*J_*E)_{M_h,F}=(E,J_*E)_{M_h,F}\).
Pairing~\eqref{eq:response-elliptic-split} with \(E\), applying
Cauchy--Schwarz to the two bounded terms, and using the nuclear
form bound yields the estimate with
\(C_\epsilon=c_\epsilon+C_V+C_K\).
\end{proof}

With exact \(L^2\) projection and integration, uniform \(L^\infty\)
bounds on the stationary orbitals and pair potentials imply the product
and multiplication bounds.
On a fixed domain, the Galerkin Coulomb map is uniformly \(L^2\)-bounded,
and Hardy's inequality~\cite{Nonnenmacher2021} gives the nuclear form bound.

Approximate Coulomb maps and quadrature must satisfy the corresponding
discrete stability bounds.
Uniform stationary-field and multiplier bounds require separate
justification for the solution family.
Supporting estimates are given in the supplementary material.

For \(E\in\mathcal T_h\setminus\{0\}\), the Rayleigh quotient
\[
\mu_h(E)=\frac{|E|_{1,h}^2}{\|E\|_h^2}
\]
measures its spatial variation.
Dividing~\eqref{eq:response-garding} by \(|E|_{1,h}^2\) gives
\begin{equation}
\label{eq:kinetic-dominance}
\frac{\left|(E,L_hE)_{M_h,F}-\tfrac12|E|_{1,h}^2\right|}
{|E|_{1,h}^2}
\leq\epsilon+\frac{C_\epsilon}{\mu_h(E)}.
\end{equation}
For sufficiently large \(\mu_h(E)\), the kinetic term dominates
the response in this quadratic-form sense.
For spatially smooth directions, Coulomb and exchange contributions
can remain significant and affect the energy curvature.
The full response need not be positive definite.

The full self-consistent RHF response thus retains an elliptic principal
part, providing the structural basis for the coupled FAS approach.


\section{Numerical Experiments}
\label{sec:numerics}

We evaluate FAS-H, the FAS method combined with a horizontal response
correction (Section~\ref{sec:physical-correction}), through three sets of
experiments. First, we examine its convergence under grid refinement.
Second, we compare its computational cost with SCF and horizontal Newton.
Finally, we use local error probes to examine the role of the multigrid
hierarchy within FAS-H.

\subsection{Experimental setup}
\label{sec:implementation}
\label{sec:structural-tests}
\label{sec:backend}

We consider eight closed-shell systems: LiH, \(\mathrm{Li_2}\),
\(\mathrm{BeH^+}\), \(\mathrm{BeH_2}\), BH, \(\mathrm{H_2O}\),
\(\mathrm{N_2}\), and \(\mathrm{CH_4}\), containing 4, 6, 4, 6, 6,
10, 14, and 10 electrons, respectively. The bond lengths of LiH,
\(\mathrm{Li_2}\), \(\mathrm{BeH^+}\), BH, and \(\mathrm{N_2}\)
are 3.015, 5.05, 2.50, 2.336, and 2.10 bohr, respectively. For linear
\(\mathrm{BeH_2}\), Be is placed at the origin and the H atoms at
\((\pm2.54,0,0)\) bohr. For \(\mathrm{H_2O}\), the O--H bond length
is 1.8088458466 bohr and the bond angle is \(104.5^\circ\). For
\(\mathrm{CH_4}\), C is placed at the origin and the four H atoms at
\(2.06/\sqrt{3}\) times \((1,1,1)\), \((1,-1,-1)\), \((-1,1,-1)\),
and \((-1,-1,1)\) bohr. All geometries are fixed; no geometry optimization
is performed.

All experiments use conforming \(Q_1\) elements on nested tensor-product
grids for both orbitals and pair potentials. This grid structure permits
efficient free-space Coulomb evaluation using the Vico--Greengard--Ferrando
(VGF) construction, with DCT precomputation and two voxels per cell
edge~\cite{VicoGreengardFerrando2016,MayaniEtAl2025}. The computational
domain is \(\Omega=[-8,8]^3\). Starting from \(h_0=4\) bohr, the mesh
width is halved at each level. The complete solves use levels L3, L4,
and L5, with mesh widths of \(0.5\), \(0.25\), and \(0.125\) bohr and
29\,791, 250\,047, and 2\,048\,383 degrees of freedom per occupied
orbital, respectively.

In all experiments, each FAS cycle uses the aligned component
correction~\eqref{eq:component-orbital-correction}, two pre-smoothing sweeps,
and two post-smoothing sweeps, with L0 as the coarsest level. The forced coarse
problem is solved by Newton--Krylov iteration with a residual target of
\(10^{-10}\) and a maximum of 30 iterations.

For smoothing, Gauss--Seidel trials use weights \(2^{-j}\),
\(j=0,\ldots,4\), and require a relative reduction of at least
\(10^{-5}\) in the complete forced residual. If none is accepted, a
safeguarded reduced Newton step is attempted; if that also fails to
reduce the residual, the state remains unchanged. On the coarsest level,
Newton backtracking uses weights \(2^{-j}\), \(j=0,\ldots,5\), and
requires a decrease in the complete forced residual. If no trial is
accepted, the solver attempts two Gauss--Seidel steps and stops unless
they decrease the residual.

The horizontal response correction uses a trigger threshold of
\(\theta=0.4\) and at most one Newton step per cycle. Each Newton step
uses GMRES with at most 60 iterations, restart 60, a base tolerance of
\(10^{-8}\), and a forcing ceiling of \(10^{-2}\). A trial is accepted
only if the rebuilt complete residual decreases.

The purpose of these experiments is to assess the effectiveness of the
proposed algorithm for solving the discrete RHF equations, rather than to
obtain highly accurate molecular solutions. Accordingly, the convergence
and reference errors reported below are measured with respect to the
discrete equations.

\subsection{Convergence under grid refinement}
\label{sec:refinement}

We first examine the convergence of FAS-H under grid refinement.
Table~\ref{tab:system-refinement} reports the numbers of FAS cycles and
accepted horizontal response steps for eight systems on L3, L4, and L5.
On each grid, the initial state is obtained by solving on coarser grids
and prolongating the resulting orbitals to the target grid.

Each solve is stopped when the complete residual falls below \(10^{-8}\).
The resulting solution is also checked against an independently computed
reference for the same discrete equations.

\begin{table}[htbp]
\centering\footnotesize\setlength{\tabcolsep}{3pt}
\caption{Convergence of FAS-H on three grid levels. Each entry $k/m$ gives the number of FAS cycles $k$ and the number of accepted horizontal response steps $m$.}\label{tab:system-refinement}
\begin{tabular}{lrrrr}\toprule
System & $N_e$ & L3 $k/m$ & L4 $k/m$ & L5 $k/m$\\\midrule
LiH & 4 & 7/0 & 7/0 & 8/0\\
Li$_2$ & 6 & 7/0 & 7/0 & 8/0\\
BeH$^+$ & 4 & 10/0 & 10/0 & 10/0\\
BeH$_2$ & 6 & 7/0 & 7/0 & 8/0\\
BH & 6 & 6/3 & 7/1 & 6/1\\
H$_2$O & 10 & 8/0 & 4/3 & 7/2\\
N$_2$ & 14 & 11/0 & 10/1 & 9/1\\
CH$_4$ & 10 & 7/0 & 8/0 & 10/0\\
\bottomrule\end{tabular}\end{table}

Table~\ref{tab:system-refinement} shows that FAS-H converges in 4--11
cycles across the eight systems and three grid levels. For each system,
the cycle counts remain comparable as the grid is refined. The horizontal
response correction is accepted in some cases, while the remaining solves
converge using FAS cycles alone.

\subsection{Comparison with Newton and SCF}
\label{sec:performance}

We compare the computational cost of FAS-H with that of SCF and
horizontal Newton on the eight systems at levels L3, L4, and L5. Here,
horizontal Newton applies the horizontal Newton step described in
Section~\ref{sec:physical-correction} as a standalone nonlinear solver.
For each system and grid, all three methods start from the same initial
state and use the same reference solution. For each comparison, the runs use eight CPUs on the same compute node.
The L5 ratios use qualified same-node reruns where needed; some measurements
were made in separate allocations.

We compare numerical solve time \(T\) and weighted FFT work
\(\mathcal W\). The weighted work expresses Coulomb evaluations in units
of a finest-grid convolution.
Specifically, let \(n_\ell\) be the number of scalar Coulomb sources
evaluated on level \(\ell\) after symmetry packing, and let \(P_\ell\)
be the size of its padded FFT grid; we define
\begin{equation}
\label{eq:fft-work}
\mathcal W=\sum_{\ell=0}^{L}n_\ell
\frac{P_\ell\log_2P_\ell}{P_L\log_2P_L}.
\end{equation}

\begin{figure}[htbp]
\centering
\includegraphics[width=\linewidth]{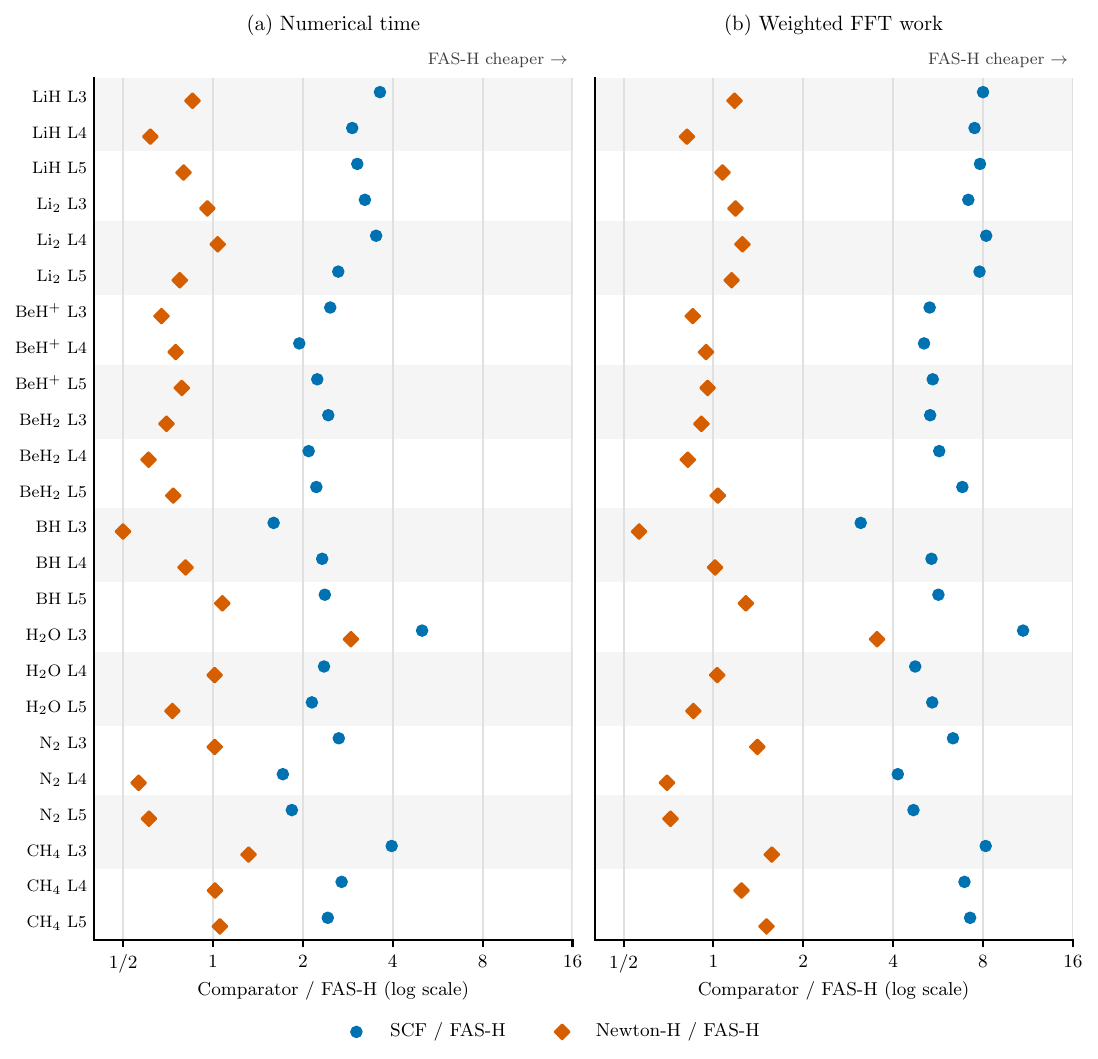}
\caption{Computational cost relative to FAS-H on 24 system--grid combinations
(sixteen on L3/L4 and eight on L5): (a) numerical time and (b) weighted FFT work.
Circles and diamonds represent SCF and horizontal Newton, respectively.
Values greater than one indicate that FAS-H has lower cost.}
\label{fig:current-cost-ratios}
\end{figure}

Figure~\ref{fig:current-cost-ratios} compares numerical time and weighted
FFT work for the three methods.

FAS-H is faster and uses less weighted FFT work than SCF in all 24
cases. SCF takes 1.60--5.02 times as long, while FAS-H reduces weighted
work by 67.9--90.8\%.

The comparison with horizontal Newton depends on the system and the cost
measure. FAS-H is faster in eight cases and uses less weighted FFT work in
fourteen. For example, on H$_2$O L3, FAS-H uses 861.8 units of weighted work
and takes 120.5 s, compared with 3045.0 units and 349.1 s for horizontal
Newton. On N$_2$ L4, the ordering is reversed: FAS-H uses 3645.4 units
and takes 4690.5 s, compared with 2548.0 units and 2642.1 s for horizontal
Newton.

FAS-H consistently outperforms SCF in the tested cases, but neither FAS-H
nor standalone horizontal Newton is uniformly faster. Because FAS-H also
uses horizontal Newton corrections, this comparison does not isolate the
contribution of the multilevel cycles.

\subsection{Multigrid mechanism}
\label{sec:mechanism}

To illustrate the effect of the multigrid cycle, we replay the first FAS
cycle for each of the eight L4 solves in Table~\ref{tab:system-refinement}.
Starting from the initial state, we track the orbital error after
pre-smoothing, coarse-grid correction, and post-smoothing.

\begin{figure}[htbp]
\centering
\includegraphics[width=\linewidth]{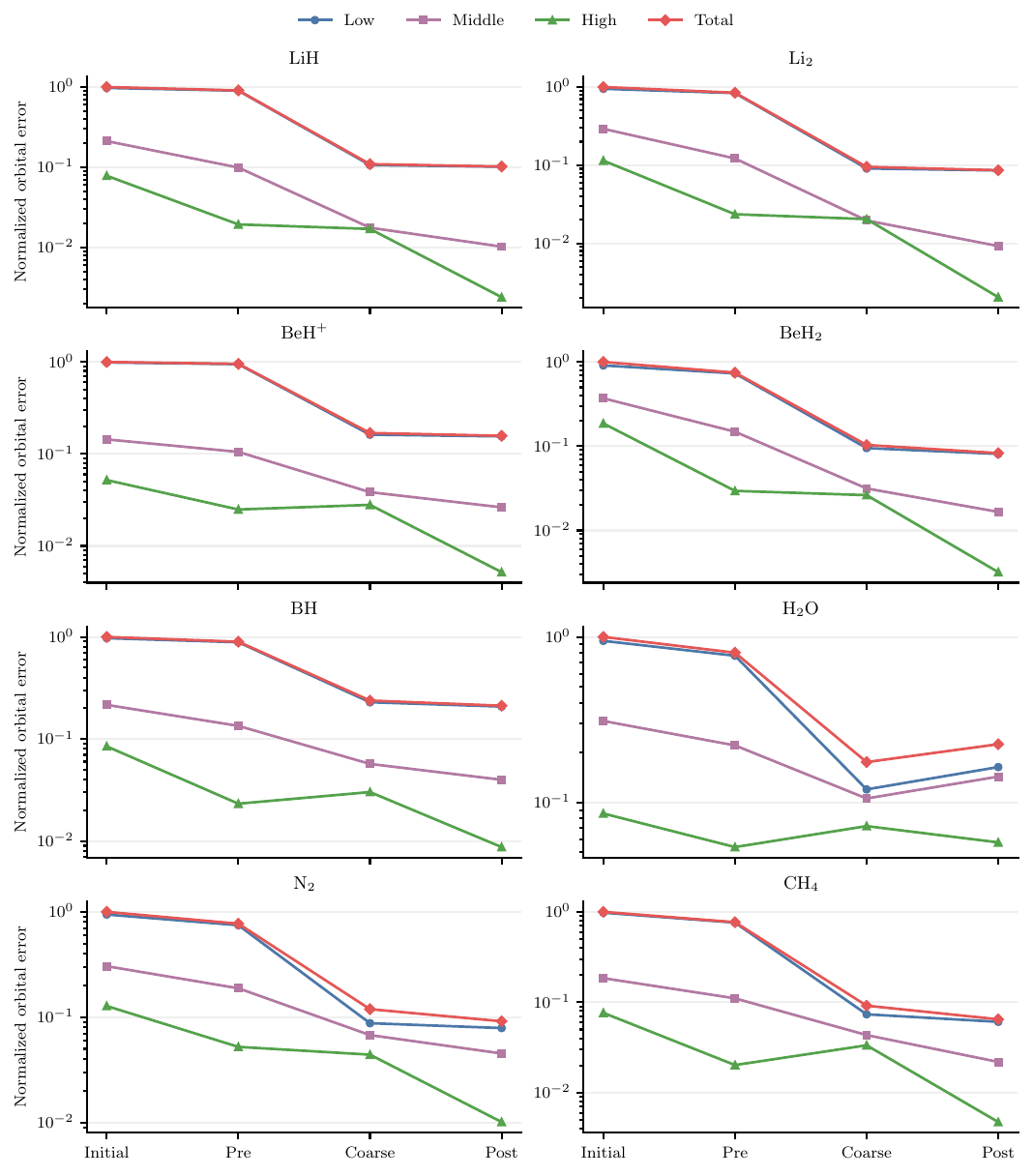}
\caption{Orbital error during the first FAS cycle for eight systems on L4.
All error norms are normalized by the initial total orbital error.
The low- and high-frequency bands contain sine modes with $\max_j k_j\leq N/4$
and $\max_j k_j>N/2$, respectively, where $N$ is the number of grid
cells per edge; the middle band contains the remaining modes. The subsequent
horizontal response correction triggered for H$_2$O is not included.}
\label{fig:fas-h-phases}
\end{figure}

The orbital error is measured relative to a discrete reference solution
after alignment of the occupied orbitals and projection onto the horizontal
space. Figure~\ref{fig:fas-h-phases} shows that pre-smoothing reduces the
high-frequency error, while the coarse-grid correction substantially reduces
the low-frequency error in all eight cases. Post-smoothing further reduces
the high-frequency error; for H$_2$O, the low-frequency error increases during
post-smoothing.

\section{Conclusion and Future Work}
\label{sec:conclusion}
We developed a structure-preserving FAS method for the complete coupled
real-space RHF system. The method treats the orbital equations, pair
potentials, and orthonormality constraints together across a hierarchy of
finite-element spaces. Forced Gram targets, occupied-frame alignment, and
auxiliary-variable reconstruction provide constraint compatibility and
occupied-rotation equivariance under the stated assumptions. Analysis of the
full self-consistent RHF response identifies an elliptic kinetic principal
part, providing a structural justification for the multilevel approach. 

Numerical experiments demonstrate convergence in a comparable number of FAS-H 
cycles across three grid levels and lower computational cost than SCF in the 
tested cases. The comparison with standalone horizontal Newton is mixed, 
indicating that the benefit of the multilevel correction depends on the problem 
and the relative costs of coarse-grid and Newton updates. 
The results support the feasibility of applying FAS directly to the coupled RHF 
system, but do not establish uniform convergence, nor superiority over 
Newton-based solvers.

Two directions may improve the efficiency of the present method. First,
interpolative separable density fitting (ISDF) can be used to compress
orbital products and reduce pair-field storage~\cite{LuYing2015,HuLinYang2017},
while the adaptively compressed exchange (ACE) approach offers a way to
reduce the cost of exchange applications~\cite{Lin2016ACE}. Second,
structured quasi-Newton methods for optimization with orthogonality
constraints~\cite{HuJiangLinWenYuan2019} may provide an efficient
approximation to the RHF response, combining a structured treatment of the
known part with a limited-memory approximation of the remaining part.

Adaptive mesh refinement is another natural extension. Refining the mesh
near nuclei and in regions where the orbitals vary rapidly could improve
spatial resolution while limiting the number of degrees of freedom.
Combining adaptivity with the coupled FAS framework would require suitable
transfer operators for the orbital and pair-potential spaces across locally
refined meshes.

The coupled FAS framework may also be extended to other electronic-structure
models. Hybrid Kohn--Sham models combine nonlocal exchange with an
exchange--correlation potential, while multiconfiguration Hartree--Fock
(MCHF) introduces a multiconfiguration wave function with coupled orbital and
configuration coefficients. These extensions would broaden the range of
electronic states and correlation effects accessible to the multilevel
approach.

\section*{Code availability}
The source code and instructions for reproducing the numerical experiments
are available at \url{https://github.com/ninotreve/FAS-for-RHF}.

\section*{Acknowledgments}
OpenAI's ChatGPT and Codex were used to assist with mathematical discussions, software implementation, numerical experimentation, and manuscript preparation. The author reviewed and validated the mathematical arguments and computational results and takes full responsibility for the content of this work.

\appendix
\section{Constrained energy curvature}
\label{app:constrained-curvature}

Let $\Phi_*$ be the discrete stationary frame of Section~\ref{sec:schur-response}.
Assume reciprocal $\mathcal C_h$ in the pair mass inner product and
consistent weak products. Let $h_0=\tfrac12A_h+V_{{\rm nuc},h}$ be
mass-self-adjoint, and define the discrete electronic energy by
\begin{equation}
E_h(\Phi)=2\sum_i(\phi_i,h_0\phi_i)_{V,h}
 +\sum_{i,j}\left[2(s_{ii},\mathcal C_hs_{jj})_{Q,h}
                    -(s_{ij},\mathcal C_hs_{ij})_{Q,h}\right],
\end{equation}
where $s_{ij}=\Pi_h(\phi_i\phi_j)$. Differentiation and reciprocity give
\[
\begin{aligned}
DE_h(\Phi)[E]
 &=4\sum_i(e_i,h_0\phi_i)_{V,h}
   +8\sum_{i,j}(\Pi_h(e_i\phi_i),\mathcal C_hs_{jj})_{Q,h}\\
 &\quad-4\sum_{i,j}(\Pi_h(e_i\phi_j),\mathcal C_hs_{ij})_{Q,h}\\
 &=4(E,\cH_h(\Phi,\mathcal C_hs_h(\Phi)))_{M_h,F}.
\end{aligned}
\]
The weak adjoints are essential for the last equality with consistent
finite-element masses. For
$\mathcal L=E_h-2\operatorname{tr}[\Lambda_*(G_h-I)]$, the orbital
gradient is $4[\cH_h(\Phi,\mathcal C_hs_h(\Phi))-\Phi\Lambda_*]$,
whose derivative at $\Phi_*$ is $4J_*$. On
$\mathcal T_h=\{E:\Phi_*^{\mathsf T}M_hE=0\}$, the mass-orthogonal
projection $P_*=I-\Phi_*\Phi_*^{\mathsf T}M_h$ removes multiplier
variations. Therefore $L_h=P_*J_*|_{\mathcal T_h}$ satisfies
\[
D^2\mathcal L(\Phi_*)[E,E]=4(E,L_hE)_{M_h,F}.
\]
Symmetry of the second variation gives mass-self-adjointness on
$\mathcal T_h$. Positive definiteness there is sufficient for a strict
local minimum modulo occupied rotations in the represented directions;
negative curvature establishes local instability.

\bibliographystyle{plain}
\bibliography{references}

@Article{Lin2016ACE,
  author = {Lin, Lin},
  title = {Adaptively Compressed Exchange Operator},
  journal = {J. Chem. Theory Comput.},
  volume = {12},
  number = {5},
  pages = {2242--2249},
  year = {2016},
  doi = {10.1021/acs.jctc.6b00092}
}

@Article{LuYing2015,
  author = {Lu, Jianfeng and Ying, Lexing},
  title = {Compression of the Electron Repulsion Integral Tensor in
           Tensor Hypercontraction Format with Cubic Scaling Cost},
  journal = {J. Comput. Phys.},
  volume = {302},
  pages = {329--335},
  year = {2015},
  doi = {10.1016/j.jcp.2015.09.014}
}

@Article{HuLinYang2017,
  author = {Hu, Wei and Lin, Lin and Yang, Chao},
  title = {Interpolative Separable Density Fitting Decomposition for
           Accelerating Hybrid Density Functional Calculations with
           Applications to Defects in Silicon},
  journal = {J. Chem. Theory Comput.},
  volume = {13},
  number = {11},
  pages = {5420--5431},
  year = {2017},
  doi = {10.1021/acs.jctc.7b00807}
}

@Article{Boys1950,
  author = {Boys, S. F.},
  title = {Electronic Wave Functions. {I}. A General Method of Calculation
           for the Stationary States of Any Molecular System},
  journal = {Proc. Roy. Soc. London Ser. A},
  volume = {200},
  number = {1063},
  pages = {542--554},
  year = {1950},
  doi = {10.1098/rspa.1950.0036}
}

@Book{HelgakerJorgensenOlsen2000,
  author = {Helgaker, Trygve and J{\o}rgensen, Poul and Olsen, Jeppe},
  title = {Molecular Electronic-Structure Theory},
  publisher = {John Wiley \& Sons},
  year = {2000},
  doi = {10.1002/9781119019572}
}

@Misc{Nonnenmacher2021,
  author = {Nonnenmacher, St{\'e}phane},
  title = {Introduction to Spectral Theory},
  year = {2021},
  howpublished = {Lecture notes, Universit{\'e} Paris-Saclay},
  note = {Proposition 3.2.7 (Hardy inequality)},
  url = {https://www.imo.universite-paris-saclay.fr/~nonnenma/enseign/Spectral%20Theory%202021%20chap1-4.pdf}
}

@Article{Roothaan1951,
  author = {Roothaan, Clemens C. J.},
  title = {New Developments in Molecular Orbital Theory},
  journal = {Rev. Modern Phys.},
  volume = {23},
  number = {2},
  pages = {69--89},
  year = {1951},
  doi = {10.1103/RevModPhys.23.69}
}

@Article{Beck2000,
  author = {Beck, Thomas L.},
  title = {Real-Space Mesh Techniques in Density-Functional Theory},
  journal = {Rev. Modern Phys.},
  volume = {72},
  number = {4},
  pages = {1041--1080},
  year = {2000},
  doi = {10.1103/RevModPhys.72.1041}
}

@Article{Davstad1992,
  author = {Davstad, Kjell},
  title = {A Multigrid Conjugate Residual Method for the Numerical Solution of
           the {Hartree--Fock} Equation for Diatomic Molecules},
  journal = {J. Comput. Phys.},
  volume = {99},
  number = {1},
  pages = {33--38},
  year = {1992}
}

@Article{CostinerTaasan1995,
  author = {Costiner, Sorin and Ta'asan, Shlomo},
  title = {Simultaneous Multigrid Techniques for Nonlinear Eigenvalue Problems:
           Solutions of the Nonlinear {Schr\"odinger--Poisson} Eigenvalue
           Problem in Two and Three Dimensions},
  journal = {Phys. Rev. E},
  volume = {52},
  number = {1},
  pages = {1181--1192},
  year = {1995},
  doi = {10.1103/PhysRevE.52.1181}
}

@Article{WangBeck2000,
  author = {Wang, Jian and Beck, Thomas L.},
  title = {Efficient Real-Space Solution of the {Kohn--Sham} Equations with
           Multiscale Techniques},
  journal = {J. Chem. Phys.},
  volume = {112},
  number = {21},
  pages = {9223--9228},
  year = {2000},
  doi = {10.1063/1.481543}
}

@Article{WijesekeraFengBeck2003,
  author = {Wijesekera, Nimal and Feng, Guogang and Beck, Thomas L.},
  title = {Multiscale Algorithms for Eigenvalue Problems},
  journal = {J. Theor. Comput. Chem.},
  volume = {2},
  number = {4},
  year = {2003},
  doi = {10.1142/S0219633603000665}
}

@Misc{BeckHeinemannKolb2003,
  author = {Beck, O. and Heinemann, D. and Kolb, D.},
  title = {Fast and Accurate Molecular {Hartree--Fock} with a Finite-Element
           Multigrid Method},
  year = {2003},
  eprint = {physics/0307108},
  archivePrefix = {arXiv}
}

@Article{Kummel2004,
  author = {K{\"u}mmel, Stephan},
  title = {Damped Gradient Iteration and Multigrid Relaxation: Tools for
           Electronic Structure Calculations Using Orbital Density-Functionals},
  journal = {J. Comput. Phys.},
  volume = {201},
  number = {1},
  pages = {333--343},
  year = {2004},
  doi = {10.1016/j.jcp.2004.05.018}
}

@Article{FengBeck2006,
  author = {Feng, Guogang and Beck, Thomas L.},
  title = {Nonlinear Multigrid Eigenvalue Solver Utilizing Nonorthogonal
           Localized Orbitals},
  journal = {Phys. Status Solidi B},
  volume = {243},
  number = {5},
  pages = {1054--1062},
  year = {2006},
  doi = {10.1002/pssb.200541446}
}

@Article{WijesekeraFengBeck2007,
  author = {Wijesekera, Nimal R. and Feng, Guogang and Beck, Thomas L.},
  title = {Efficient Multiscale Algorithms for Solution of Self-Consistent
           Eigenvalue Problems in Real Space},
  journal = {Phys. Rev. B},
  volume = {75},
  number = {11},
  pages = {115101},
  year = {2007},
  doi = {10.1103/PhysRevB.75.115101}
}

@Article{AlizadeganHsiaMartinez2010,
  author = {Alizadegan, Reza and Hsia, K. Jimmy and Martinez, Todd J.},
  title = {A Divide and Conquer Real Space Finite-Element {Hartree--Fock}
           Method},
  journal = {J. Chem. Phys.},
  volume = {132},
  number = {3},
  pages = {034101},
  year = {2010},
  doi = {10.1063/1.3290949}
}

@Article{KhoromskijKhoromskaiaFlad2011,
  author = {Khoromskij, Boris N. and Khoromskaia, Venera and Flad, Heinz-Juergen},
  title = {Numerical Solution of the {Hartree--Fock} Equation in Multilevel
           Tensor-Structured Format},
  journal = {SIAM J. Sci. Comput.},
  volume = {33},
  number = {1},
  pages = {45--65},
  year = {2011},
  doi = {10.1137/090777372}
}

@Article{Braun2014,
  author = {Braun, Moritz},
  title = {Finite Element {Hartree--Fock} Calculations in Three Dimensions for
           Atoms and Small Molecules},
  journal = {J. Comput. Appl. Math.},
  volume = {270},
  pages = {100--108},
  year = {2014},
  doi = {10.1016/j.cam.2014.02.012}
}

@Article{HaubenwallnerHeller2025,
  author = {Haubenwallner, Paul and Heller, Matthias},
  title = {Fully Numerical {Hartree--Fock} Calculations for Atoms and Small
           Molecules with Quantics Tensor Trains},
  journal = {Electron. Struct.},
  volume = {7},
  number = {2},
  pages = {025006},
  year = {2025},
  doi = {10.1088/2516-1075/add9f8}
}

@Article{BoffiJainNatan2016,
  author = {Boffi, Nicholas M. and Jain, Manish and Natan, Amir},
  title = {Efficient Computation of the {Hartree--Fock} Exchange in Real-Space
           with Projection Operators},
  journal = {J. Chem. Theory Comput.},
  volume = {12},
  number = {8},
  pages = {3614--3622},
  year = {2016},
  doi = {10.1021/acs.jctc.6b00376}
}

@Article{MRChem2023,
  author = {Wind, Peter and Bj{\o}rgve, Magnar and Brakestad, Anders
            and {Gerez S.}, Gabriel A. and Jensen, Stig Rune
            and {Di Remigio Eik{\aa}s}, Roberto and Frediani, Luca},
  title = {{MRChem} Multiresolution Analysis Code for Molecular Electronic
           Structure Calculations: Performance and Scaling Properties},
  journal = {J. Chem. Theory Comput.},
  volume = {19},
  number = {1},
  pages = {137--146},
  year = {2023},
  doi = {10.1021/acs.jctc.2c00982}
}

@Misc{PanLindsey2025,
  author = {Pan, Yulong and Lindsey, Michael},
  title = {Fast Adaptive Discontinuous Basis Sets for Electronic Structure},
  year = {2025},
  eprint = {2510.21213},
  archivePrefix = {arXiv},
  primaryClass = {physics.comp-ph},
  doi = {10.48550/arXiv.2510.21213}
}

@Article{XuLiLiang2026,
  author = {Xu, Fei and Li, Yuting and Liang, Lu},
  title = {Multilevel Correction Adaptive Finite Element Method for
           {Hartree--Fock} Equation},
  journal = {J. Chem. Phys.},
  volume = {164},
  number = {7},
  pages = {074111},
  year = {2026},
  doi = {10.1063/5.0310981}
}

@Article{XuLiangLi2026,
  author = {Xu, Fei and Liang, Lu and Li, Yuting},
  title = {An Enhanced Adaptive Finite Element Method with Linear
           Scaling for {3D} {Hartree--Fock} Equation},
  journal = {J. Chem. Phys.},
  volume = {165},
  pages = {164109},
  month = oct,
  year = {2026},
  doi = {10.1063/5.0356545}
}

@Article{Brandt1977,
  author = {Brandt, Achi},
  title = {Multi-Level Adaptive Solutions to Boundary-Value Problems},
  journal = {Math. Comp.},
  volume = {31},
  number = {138},
  pages = {333--390},
  year = {1977},
  doi = {10.1090/S0025-5718-1977-0431719-X}
}

@Article{Pulay1982,
  author = {Pulay, Peter},
  title = {Improved {SCF} Convergence Acceleration},
  journal = {J. Comput. Chem.},
  volume = {3},
  number = {4},
  pages = {556--560},
  year = {1982},
  doi = {10.1002/jcc.540030413}
}

@Article{KudinScuseriaCances2002,
  author = {Kudin, Konstantin N. and Scuseria, Gustavo E. and Canc{\`e}s, Eric},
  title = {A Black-Box Self-Consistent Field Convergence Algorithm: One Step Closer},
  journal = {J. Chem. Phys.},
  volume = {116},
  number = {19},
  pages = {8255--8261},
  year = {2002},
  doi = {10.1063/1.1470195}
}

@article{VanVoorhisHeadGordon2002,
  author = {Van Voorhis, Troy and Head-Gordon, Martin},
  title = {A geometric approach to direct minimization},
  journal = {Mol. Phys.},
  volume = {100},
  number = {11},
  pages = {1713--1721},
  year = {2002},
  doi = {10.1080/00268970110103642}
}

@article{HuJiangLinWenYuan2019,
  author = {Hu, Jiang and Jiang, Bo and Lin, Lin and Wen, Zaiwen and Yuan, Yaxiang},
  title = {Structured Quasi-{Newton} Methods for Optimization with Orthogonality Constraints},
  journal = {SIAM J. Sci. Comput.},
  volume = {41},
  number = {4},
  pages = {A2239--A2269},
  year = {2019},
  doi = {10.1137/18M121112X},
  eprint = {1809.00452},
  archivePrefix = {arXiv}
}

@misc{Dinvay2026,
  author = {Dinvay, Evgueni},
  title = {Riemannian gradient descent for {Hartree--Fock} theory},
  year = {2026},
  eprint = {2603.15870},
  archivePrefix = {arXiv},
  primaryClass = {quant-ph},
  note = {Preprint}
}

@article{VicoGreengardFerrando2016,
  author = {Vico, Felipe and Greengard, Leslie and Ferrando, Miguel},
  title = {Fast convolution with free-space {Green}'s functions},
  journal = {J. Comput. Phys.},
  volume = {323},
  pages = {191--203},
  year = {2016},
  doi = {10.1016/j.jcp.2016.07.028},
  eprint = {1604.03155},
  archivePrefix = {arXiv}
}

@misc{MayaniEtAl2025,
  author = {Mayani, Sonali and Montanaro, Veronica and Cerfon, Antoine and
            Frey, Matthias and Muralikrishnan, Sriramkrishnan and Adelmann, Andreas},
  title = {A Massively Parallel Performance Portable Free-space Spectral {Poisson} Solver},
  year = {2025},
  note = {Preprint, version 2},
  url = {https://arxiv.org/abs/2405.02603v2}
}

\end{document}


\maketitle

\section{Coupled response and supporting estimates}

This supplement supplies the algebra and stability estimates used in
Section~4 of the main paper. All operators and pairings are those of the
discrete model in Section~3. The analysis concerns the represented
discrete equations on a fixed bounded domain $\Omega\subset\R^3$.

\subsection{Product adjoints and the coupled Jacobian}

For orbital and pair mass matrices $M_h$ and $M_{Q,h}$, define
\[
\mathcal T_{h,u}v=\Pi_h(uv),\qquad
\mathcal T_{h,u}^{\dagger}
=M_h^{-1}\mathcal T_{h,u}^{\mathsf T}M_{Q,h}.
\]
The weak multiplication convention gives
$\mathcal M_{h,w}u=\mathcal T_{h,u}^{\dagger}w$ and
$(v,\mathcal M_{h,w}u)_{V,h}=(\Pi_h(uv),w)_{Q,h}$.
For fixed $w$, $\mathcal M_{h,w}$ is mass-self-adjoint.

Let $X_*=(\Phi_*,W_*,\Lambda_*)$ be an unforced stationary state with
$\Phi_*^{\mathsf T}M_h\Phi_*=I$, $W_*=\mathcal C_hs_h(\Phi_*)$,
and symmetric $\Lambda_*$. Keep $\Pi_h$, $\mathcal C_h$, and
$\mathcal D_h$ fixed and linear, with $\mathcal D_h$ invertible. Set
\[
\begin{aligned}
\mathcal A_*E&=\cH_h(E,W_*)-E\Lambda_*,\\
(\mathcal B_*\dot W)_i
 &=2\sum_j\mathcal M_{h,\dot W_{jj}}\phi_{*,i}
   -\sum_j\mathcal M_{h,\dot W_{ij}}\phi_{*,j},\\
(\mathcal S_*E)_{ij}
 &=\Pi_h(e_i\phi_{*,j}+\phi_{*,i}e_j),\\
\mathcal G_*E
 &=\Phi_*^{\mathsf T}M_hE+E^{\mathsf T}M_h\Phi_*.
\end{aligned}
\]
For symmetric perturbations $\dot W$ and $\dot\Lambda$, the complete
three-block derivative is
\begin{equation}
DF_h(X_*)
\begin{pmatrix}E\\\dot W\\\dot\Lambda\end{pmatrix}
=
\begin{pmatrix}
\mathcal A_*E+\mathcal B_*\dot W-\Phi_*\dot\Lambda\\
\mathcal D_h(\dot W-\mathcal C_h\mathcal S_*E)\\
\mathcal G_*E
\end{pmatrix}.
\end{equation}
At fixed pair forcing, eliminating $\dot W$ gives
$\dot W=\mathcal C_h\mathcal S_*E$ and
$J_*=\mathcal A_*+\mathcal B_*\mathcal C_h\mathcal S_*$.

\subsection{Bounds on the interaction response}

The decomposition in the main paper has the explicit terms
\begin{equation}
\begin{aligned}
(\mathcal V_hE)_i
 &=2\sum_j\mathcal M_{h,W_{*,jj}}e_i
   -\sum_j\mathcal M_{h,W_{*,ij}}e_j-\sum_je_j\Lambda_{*,ji},\\
(\mathcal K_hE)_i
 &=4\mathcal T_{h,\phi_{*,i}}^{\dagger}
       \sum_j\mathcal C_h\mathcal T_{h,\phi_{*,j}}e_j\\
 &\quad-\sum_j\mathcal T_{h,\phi_{*,j}}^{\dagger}\mathcal C_h\Pi_h
       (\phi_{*,j}e_i+\phi_{*,i}e_j).
\end{aligned}
\end{equation}
Suppose the bounds in the main paper hold with constants
$b,w,\lambda,c_C$, so that
\[
\begin{gathered}
\max_i\|\phi_{*,i}\|_\infty\le b,\qquad
\max_{i,j}\|W_{*,ij}\|_\infty\le w,\qquad
\|\Lambda_*\|_2\le\lambda,\qquad
\|\mathcal C_h\|_{Q,h\to Q,h}\le c_C,\\
\|\mathcal T_{h,\phi_{*,i}}\|_{V,h\to Q,h}\le b,\qquad
\|\mathcal M_{h,W_{*,ij}}\|_{V,h\to V,h}\le w.
\end{gathered}
\]
Exact $L^2$ projection and integration imply the last two bounds from
the field bounds. For approximate forms, uniform stability constants
must be included. Since a mass adjoint has the same operator norm,
each product--Coulomb--adjoint term is bounded by $b^2c_C$. Summing
over occupied indices and using Cauchy--Schwarz yields
\begin{equation}
\|\mathcal V_hE\|_h\le(3Nw+\lambda)\|E\|_h,\qquad
\|\mathcal K_hE\|_h\le6Nb^2c_C\|E\|_h.
\end{equation}
Combining these bounds with the nuclear form estimate below gives
the main paper's G\aa rding estimate with
$C_\epsilon=c_\epsilon+3Nw+\lambda+6Nb^2c_C$.

\subsection{Coulomb and nuclear form bounds}

On a bounded three-dimensional domain, the kernel $|x-y|^{-1}$ is
square integrable on $\Omega\times\Omega$, so the continuum Coulomb
operator $\mathcal C_\Omega$ is bounded on $L^2(\Omega)$. For the exact
Galerkin map, orthogonal projection gives
\[
\mathcal C_h=P_{Q,h}\mathcal C_\Omega|_{Q_h},\qquad
\|\mathcal C_h\|_{Q,h\to Q,h}
\le\|\mathcal C_\Omega\|_{L^2\to L^2}.
\]
Similarly, inserting an orthogonal source projection on both sides of
the continuum Coulomb form retains this bound. Quadrature or compression
errors require an additional uniform mass-norm bound. 
For $u\in H_0^1(\Omega)$ extended by zero, Hardy's inequality in
$\R^3$ and Cauchy--Schwarz give, for each nucleus $R_a$,
\[
\int_\Omega\frac{|u(x)|^2}{|x-R_a|}\,dx
\le\|u\|_{L^2}\left\|\frac{u}{|x-R_a|}\right\|_{L^2}
\le2\|u\|_{L^2}\|\nabla u\|_{L^2}.
\]
For a fixed finite set of nuclei, summation and Young's inequality
therefore imply
\[
\bigl|(E,V_{{\rm nuc},h}E)_{M_h,F}\bigr|
\le\epsilon|E|_{1,h}^2+c_\epsilon\|E\|_h^2
\]
under exact conforming integration. With numerical nuclear quadrature,
the corresponding uniform discrete estimate is an additional assumption.
\section{Data for the figures}

Table~\ref{tab:s-cost-ratios} gives the values plotted in the cost-ratio
figure in the main paper. Each cost is divided by the corresponding FAS-H
cost. The L5 comparisons use same-node measurements, including reruns;
some measurements were made in separate allocations.

\begin{table}[htbp]
\centering\footnotesize
\caption{Costs relative to FAS-H. $T$ is numerical solve time and
$\mathcal W$ is weighted FFT work; $S$ and $H$ denote SCF and horizontal
Newton, respectively.}\label{tab:s-cost-ratios}
\begin{tabular}{llrrrr}\toprule
System & Level & $T_S/T_F$ & $T_H/T_F$ & $\mathcal W_S/\mathcal W_F$ & $\mathcal W_H/\mathcal W_F$\\\midrule
LiH & L3 & 3.6238 & 0.85322 & 8.001 & 1.1774\\
LiH & L4 & 2.9258 & 0.61682 & 7.4928 & 0.8154\\
LiH & L5 & 3.0423 & 0.7966 & 7.8159 & 1.0726\\
Li$_2$ & L3 & 3.2261 & 0.95727 & 7.1425 & 1.1856\\
Li$_2$ & L4 & 3.5178 & 1.0367 & 8.1971 & 1.2511\\
Li$_2$ & L5 & 2.6273 & 0.77331 & 7.7869 & 1.1508\\
BeH$^+$ & L3 & 2.4695 & 0.67185 & 5.3004 & 0.85278\\
BeH$^+$ & L4 & 1.9448 & 0.74944 & 5.077 & 0.94509\\
BeH$^+$ & L5 & 2.2343 & 0.78605 & 5.4275 & 0.95615\\
BeH$_2$ & L3 & 2.4324 & 0.69835 & 5.3199 & 0.91146\\
BeH$_2$ & L4 & 2.0907 & 0.6083 & 5.7065 & 0.8208\\
BeH$_2$ & L5 & 2.2184 & 0.73567 & 6.8215 & 1.0344\\
BH & L3 & 1.5958 & 0.49951 & 3.1144 & 0.56363\\
BH & L4 & 2.3203 & 0.80888 & 5.3745 & 1.0125\\
BH & L5 & 2.3681 & 1.0744 & 5.6703 & 1.2845\\
H$_2$O & L3 & 5.015 & 2.8968 & 10.896 & 3.5332\\
H$_2$O & L4 & 2.3537 & 1.0122 & 4.7432 & 1.0297\\
H$_2$O & L5 & 2.1433 & 0.73078 & 5.407 & 0.85687\\
N$_2$ & L3 & 2.6377 & 1.0132 & 6.3454 & 1.403\\
N$_2$ & L4 & 1.7134 & 0.56329 & 4.1458 & 0.69896\\
N$_2$ & L5 & 1.8366 & 0.60973 & 4.6783 & 0.71816\\
CH$_4$ & L3 & 3.9661 & 1.315 & 8.1658 & 1.5685\\
CH$_4$ & L4 & 2.6956 & 1.0153 & 6.9326 & 1.2412\\
CH$_4$ & L5 & 2.4225 & 1.0538 & 7.2415 & 1.5059\\
\bottomrule\end{tabular}\end{table}

Table~\ref{tab:s-phase-errors} gives the normalized orbital errors plotted
in the first-cycle phase figure. Each value is the square root of the band
energy divided by the initial total-error norm. The phases are initial,
after pre-smoothing, after coarse correction, and after post-smoothing.

{\footnotesize\setlength{\tabcolsep}{4pt}
\begin{longtable}{llrrrr}
\caption{Normalized L4 orbital errors during the first FAS cycle.}
\label{tab:s-phase-errors}\\
\toprule System & Phase & Low & Middle & High & Total\\\midrule\endfirsthead
\multicolumn{6}{c}{\tablename\ \thetable\ (continued)}\\
\toprule System & Phase & Low & Middle & High & Total\\\midrule\endhead
LiH & Initial & 0.9742 & 0.21165 & 0.078292 & 1\\
 & Pre-smoothing & 0.90102 & 0.099022 & 0.019351 & 0.90665\\
 & Coarse correction & 0.1065 & 0.017613 & 0.016997 & 0.10928\\
 & Post-smoothing & 0.10149 & 0.010232 & 0.0023951 & 0.10203\\
\addlinespace
Li$_2$ & Initial & 0.94956 & 0.29197 & 0.11441 & 1\\
 & Pre-smoothing & 0.83588 & 0.12191 & 0.023568 & 0.84505\\
 & Coarse correction & 0.090892 & 0.019633 & 0.020438 & 0.095208\\
 & Post-smoothing & 0.085687 & 0.0092408 & 0.0020494 & 0.086208\\
\addlinespace
BeH$^+$ & Initial & 0.98821 & 0.144 & 0.052003 & 1\\
 & Pre-smoothing & 0.94741 & 0.1053 & 0.024901 & 0.95357\\
 & Coarse correction & 0.16214 & 0.038417 & 0.02796 & 0.16896\\
 & Post-smoothing & 0.15528 & 0.026287 & 0.0051908 & 0.15758\\
\addlinespace
BeH$_2$ & Initial & 0.91003 & 0.37034 & 0.18625 & 1\\
 & Pre-smoothing & 0.7318 & 0.14857 & 0.029489 & 0.74731\\
 & Coarse correction & 0.094936 & 0.031506 & 0.026238 & 0.10341\\
 & Post-smoothing & 0.081016 & 0.016604 & 0.0031975 & 0.082762\\
\addlinespace
BH & Initial & 0.97274 & 0.21588 & 0.08468 & 1\\
 & Pre-smoothing & 0.88717 & 0.13442 & 0.02323 & 0.8976\\
 & Coarse correction & 0.22918 & 0.056992 & 0.030194 & 0.23808\\
 & Post-smoothing & 0.20782 & 0.039845 & 0.0087266 & 0.21178\\
\addlinespace
H$_2$O & Initial & 0.94668 & 0.31054 & 0.085734 & 1\\
 & Pre-smoothing & 0.76991 & 0.22104 & 0.053718 & 0.80281\\
 & Coarse correction & 0.1199 & 0.10564 & 0.071938 & 0.17525\\
 & Post-smoothing & 0.16355 & 0.14321 & 0.057342 & 0.22482\\
\addlinespace
N$_2$ & Initial & 0.94381 & 0.30485 & 0.12761 & 1\\
 & Pre-smoothing & 0.74736 & 0.18803 & 0.052223 & 0.77241\\
 & Coarse correction & 0.087644 & 0.067563 & 0.04405 & 0.11911\\
 & Post-smoothing & 0.078775 & 0.045206 & 0.010088 & 0.091383\\
\addlinespace
CH$_4$ & Initial & 0.9798 & 0.18482 & 0.076406 & 1\\
 & Pre-smoothing & 0.76108 & 0.11055 & 0.020213 & 0.76934\\
 & Coarse correction & 0.073582 & 0.043386 & 0.033429 & 0.091729\\
 & Post-smoothing & 0.060795 & 0.021889 & 0.0047337 & 0.064789\\
\addlinespace
\bottomrule\end{longtable}}